\documentclass[twocolumn,english,aps,prl,showpacs,superscriptaddress]{revtex4-2}
\usepackage[T1]{fontenc}
\usepackage[latin9]{inputenc}
\usepackage{geometry}
\usepackage{xcolor}
\usepackage{babel}
\usepackage{units}
\usepackage{amsmath}
\usepackage{amssymb}
\usepackage{mathtools} %Eran added this.
\usepackage{braket} %Eran added this.
\usepackage{stmaryrd}
\usepackage{graphicx}
\usepackage{wasysym}
\usepackage[bookmarks=false]{hyperref}

\makeatletter
  
\IfFileExists{lmodern.sty}{\usepackage{lmodern}}{}
\usepackage{babel}
\usepackage{orcidlink}

\usepackage{hyperref}
\def\equationautorefname~#1\null{Equation (#1)\null}

\makeatother

\begin{document}
\title{Optical protection of alkali-metal atoms from spin relaxation}

\author{Avraham Berrebi}
\affiliation{Department of Applied Physics, The Faculty of Science, The Center for Nanoscience and Nanotechnology, The Hebrew University of Jerusalem, Jerusalem 9190401, Israel.}
\author{Mark Dikopoltsev}
\affiliation{Department of Applied Physics, The Faculty of Science, The Center for Nanoscience and Nanotechnology, The Hebrew University of Jerusalem, Jerusalem 9190401, Israel.}
\affiliation{Rafael Ltd, 31021, Haifa, Israel.}
\author{Ori Katz}
\affiliation{Department of Applied Physics, The Faculty of Science, The Center for Nanoscience and Nanotechnology, The Hebrew University of Jerusalem, Jerusalem 9190401, Israel.}
\author{Or Katz~\orcidlink{0000-0001-7634-1993}}
\email{Corresponding author: or.katz@cornell.edu}
\affiliation{School of Applied and Engineering Physics, Cornell University, Ithaca, NY 14853.}

\begin{abstract}
We present an optical technique for suppressing relaxation in alkali-metal spins using a single off-resonant laser beam. The method harnesses a physical mechanism that synchronizes Larmor precession in the two hyperfine manifolds, protecting magnetic coherence from relaxation caused by spin-exchange and other hyperfine-changing collisions. We experimentally demonstrate up to a ninefold reduction in decoherence of warm cesium vapor, achieving simultaneous protection from both spin-exchange relaxation and partial depolarization from coated cell walls. The technique substantially enhances the spin precession quality factor and maintains a stable gyromagnetic ratio independent of spin polarization, even under frequent collisions. These findings offer a pathway for mitigating dominant relaxation channels in alkali-metal-based applications and experiments, particularly in anti-relaxation-coated cells.\end{abstract}
\maketitle
Collisions are a fundamental relaxation mechanism in warm spin gases. At ambient conditions, alkali-metal atoms confined in a glass cell move at speeds of hundreds of meters per second, frequently colliding with the cell walls or with each other. During these collisions, the exchange interaction between their single valence electrons correlates the spin states of the colliding atoms \cite{happer1972optical,happer1973spin,happer1977effect, mouloudakis2021spin, katz2018synchronization}. Due to the random nature of spin-exchange collisions, the ensemble is driven towards a spin-temperature distribution \cite{anderson1959n}, and in the presence of a magnetic field,  this results in relaxation of the magnetic moments,  associated with Larmor coherences \cite{happer1972optical,happer1977effect, savukov2005effects}. This process typically dominates the relaxation in ensembles of alkali-metal atoms at room temperature and above.

Two primary techniques enable suppression of this relaxation, operating in a regime known as Spin-Exchange Relaxation Free (SERF). One approach relies on operating at low magnetic fields, where the Larmor precession frequency is small compared to the spin-exchange collision rate \cite{happer1973spin,happer1977effect, kominis2003subfemtotesla, dang2010ultrahigh, ledbetter2008spin, balabas2010polarized,  kong2020measurement, jimenez2014optically, katz2013nonlinear, chalupczak2014spin, xiao2021atomic, mouloudakis2022effects,Dikopoltsev2022,korver2013suppression}, and the other on maintaining a high degree of spin polarization \cite{jau2004intense,smullin2009low,sheng2013subfemtotesla}. Systems operating in this regime are used in a variety of applications including precision sensors \cite{kominis2003subfemtotesla, dang2010ultrahigh, kornack2005nuclear, yang2019investigation, fang2014spin, savukov2017spin,jau2004intense,zhivun2019dual}, searches for new physics \cite{wang2018application,bloch2022new, vasilakis2009limits, budker2014proposal, safronova2018search, budker2022quantum, wang2022limits, padniuk2022response,bloch2023constraints,bloch2024rotating}, coupling to noble-gas spins for imaging and fundamental studies \cite{gentile2017optically, coulter1988neutron, walker1997spin, chupp2001medical, tsinovoy2022enhanced, lee2021hyperpolarised, katugampola2021frequency, shaham2022strong,katz2021coupling}, and emerging quantum information applications \cite{kong2020measurement, katz2020long, mouloudakis2021spin, katz2018light, guarrera2021spin, katz2022quantum, mouloudakis2022effects,katz2022optical}. However, these techniques often impose practical limitations, such as the stringent requirements on on magnetic field control or high spin polarization.

Here we describe a mechanism that protects warm alkali-metal ensembles from spin relaxation using an off-resonant laser beam. Unlike previous approaches, this method does not rely on high spin polarization or low magnetic fields. Instead, protection arises from synchronizing the Larmor precession in the two hyperfine manifolds via differential vector light shifts, thereby suppressing spin relaxation caused by hyperfine-state-changing processes. We experimentally demonstrate this approach in a warm cesium vapor and observe up to a ninefold suppression of the relaxation rate, along with a substantial enhancement of the spin precession quality factor.

\begin{figure*}[t]
\begin{centering}
\includegraphics[width=16cm]{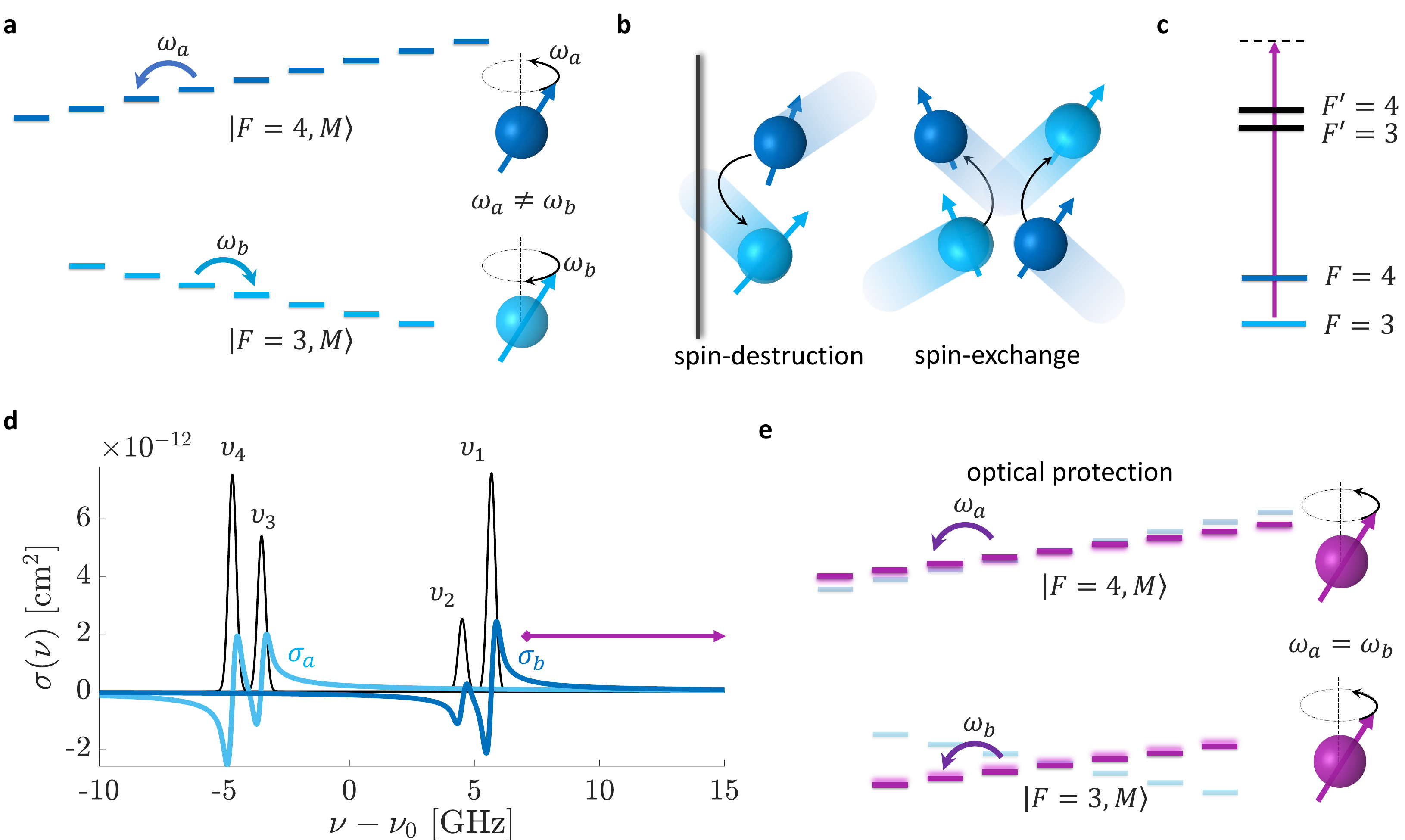}
\par\end{centering}
\centering{}\caption{\textbf{Relaxation of alkali-metal spins and light-shift compensation.} \textbf{a}, Spin levels of cesium atoms (${I=7/2}$) in the electronic ground state. The magnetic levels $\ket{F,M}$ in the $F=4$ ($F=3$)  hyperfine manifolds are Zeeman-splitted by $\hbar\omega_{\textrm{a}}$ ($\hbar\omega_{\textrm{b}}$) for $M\leq|F|$, resulting in asynchronous precession of the magnetic moments; spins precess clockwise at $F=4$ (blue) but counter-clockwise at $F=3$ (cyan). \textbf{b}, Spin-exchange collisions between pairs of atoms, or collisions with weakly-depolarizing walls of the enclosure, lead to random changes of the hyperfine manifolds and give rise to spin-relaxation by asynchronous precession. \textbf{c}, An off-resonance circularly-polarized optical beam shifts the magnetic levels within the hyperfine manifolds \textbf{d}, Calculated light-shift (complex) cross-sections $\sigma_{\textrm{a}}$ (cyan) and $\sigma_{\textrm{b}}$ (blue) of the Zeeman-like shifts in the $F=4$ and $F=3$ manifolds respectively, as a function of the optical frequency $\nu$ from the D$_1$ transitions. Black lines denote the absorption cross-section of the four transition lines and purple arrow denotes the range where the levels in $F=3$ manifold are primarily shifted. \textbf{e}, Applied Zeeman-like shifts can correct for the difference in precession frequencies and protect from dephasing by asynchronous precession. \label{fig:system}}
\end{figure*}

The asynchronous Larmor precession of magnetic moments, which underlies spin-exchange relaxation and other dephasing processes \cite{Dikopoltsev2022}, originates from the level structure of alkali-metal atoms. The hyperfine interaction between the electron spin $S=\tfrac{1}{2}$ and the nonzero nuclear spin $I$, splits the ground state into two hyperfine manifolds labeled by $F=I\pm\tfrac{1}{2}$, separated by the hyperfine frequency $\omega_{\textrm{hpf}}$. In the presence of a magnetic field, the degeneracy within each manifold is lifted, and the magnetic sublevels are labeled by the quantum number $M=S_z+I_z$, where $|M|\leq F$. The resulting Zeeman splittings are dominated by the coupling of the magnetic field to the electron spin, as shown in Fig.~\ref{fig:system}a. In the upper manifold, adjacent sublevels split by $\omega_{\textrm{a}}=+\omega_B$; in the lower manifold, the splittings are inverted, with $\omega_{\textrm{b}}=-\omega_B$, valid to leading order in $g_e B/\omega_{\textrm{hpf}}$. Here, $\omega_B=g_eB/(2I+1)$ is the Larmor precession rate in a magnetic field $B\hat{z}$ (taken to define the quantization axis along $z$), and $g_e$ is the electron gyromagnetic ratio.

This sign reversal, $\omega_{\textrm{a}}=-\omega_{\textrm{b}}$ arises from the alignment or anti-alignment of the electron and nuclear spins in the two hyperfine manifolds, resulting in opposite Larmor precession directions. At room temperature, collisions between alkali-metal atoms can induce transitions between these manifolds. Because $\omega_{\textrm{a}}\neq\omega_{\textrm{b}}$, such transitions abruptly change the precession direction, causing asynchronous dynamics and dephasing of transverse Larmor coherence, even in the case of spin-exchange collisions that conserve the total spin of the colliding pair.
 
To suppress this dephasing, we apply a circularly polarized, far-detuned "protection" beam along the magnetic field direction. The beam induces opposite vector light shifts $\delta_{\textrm{a}}$ and $\delta_{\textrm{b}}$ in the two hyperfine manifolds, while maintaining minimal photon scattering. These shifts are proportional to the beam intensity and depend on the vector light-shift cross sections $\sigma_{\textrm{a}}(\nu)$ and $\sigma_{\textrm{b}}(\nu) $, which vary with the laser detuning $\nu$ from the atomic D1 transitions (Fig.~\ref{fig:system}c-d for cesium) \cite{mathur1968light,happer1967effective,cohen1972experimental}. By tuning the detuning and intensity to satisfy the condition
\begin{equation}\label{eq:resonance_condition}
    \omega_B + \delta_{\textrm{a}} = -\omega_B + \delta_{\textrm{b}},
\end{equation}
the effective precession frequencies in both manifolds are equalized, i.e.,  $\omega_{\textrm{a}} + \delta_{\textrm{a}} = \omega_{\textrm{b}} + \delta_{\textrm{b}}$. This ensures that atoms in both hyperfine states precess in the same direction at the same rate. As a result, spin-exchange-induced transitions between them no longer alter the phase of the transverse Larmor coherence. In contrast to the SERF regime, which requires $\omega_{B}\ll R_{\textrm{se}}$ or a high degree of spin polarization to suppress relaxation, the mechanism demonstrated here enables coherent spin dynamics even at higher precession rates and low polarization. A  more detailed treatment of this protection mechanism, including an ensemble-level model of spin-exchange dynamics beyond the mean-field approximation, is provided in \cite{SI}. 

\begin{figure*}[t]
\begin{centering}
\includegraphics[width=16cm]{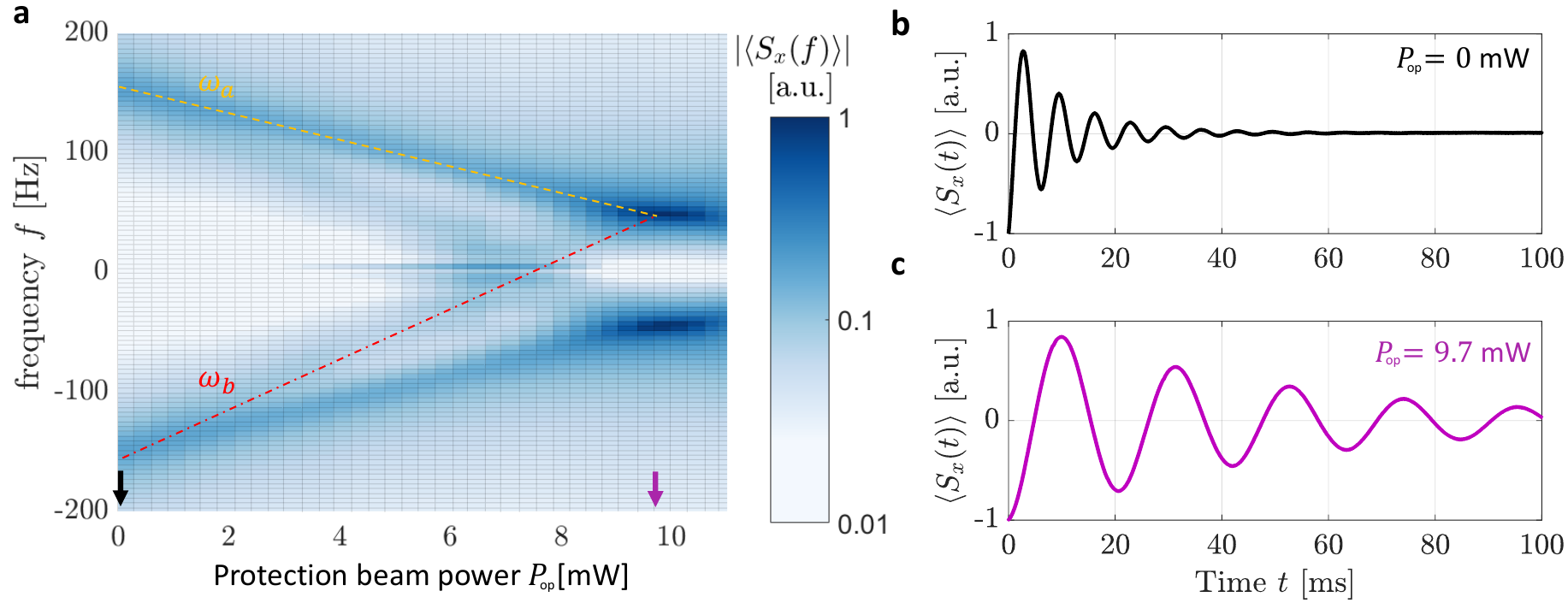}
\par\end{centering}
\centering{}\caption{\textbf{Synchronization of spin precession using light-shifts}. \textbf{a}, Fourier spectrum of the measured spin precession of cesium spins at $B=0.43\,\textrm{mG}$. The protection beam with power $P_{\textrm{op}}$ induces a Zeeman-like field that shifts the precession frequency of the magnetic moments in $F=3$ (i.e.~$\omega_{\textrm{b}}$, red dash-dots lines) and, to a lesser extent, in $F=4$ (i.e.,~$\omega_{\textrm{a}}$, orange dash line). At the resonance condition $\omega_{\textrm{a}}=\omega_{\textrm{b}}$ near $P_{\textrm{op}}=9.7\,\textrm{mW}$ the frequencies synchronize and the spectrum is significantly enhanced. \textbf{b}, Spin precession without the optical protection beam (top) and in the presence of the protection beam (bottom). At synchronized precession, the coherence time is prolonged fivefold.\label{fig:spectrum}}
\end{figure*}

We study this mechanism using a warm cesium vapor ($S=1/2,\,I=7/2$) in a buffer-gas-free, paraffin-coated cylindrical glass cell, with full details in \cite{SI}. The cesium vapor density, $n=(2.5\pm0.4)\times 10^{11} \,\textrm{cm}^{-3}$, is controlled by heating a liquid cesium droplet and measured via absorption spectroscopy \cite{SI}. Magnetic fields are controlled using Helmholtz coils and magnetic shielding. Spins are initially polarized with a weak circularly polarized optical pump beam, detuned near the $\nu_3$ transition, propagating along the $z$-axis of the cylinder with a background magnetic field applied in the $xz$ plane during pumping. Afterward, both the pump and background fields are turned off, and a magnetic field $B$ and a circularly polarized protection beam, $12$ GHz blue-detuned from the $\nu_1$ transition, are applied along $z$. The protection beam, nearly uniform in power across the cell diameter \cite{SI}, selectively induces different vector light-shifts in the two hyperfine manifolds. Spin precession is monitored using a weak, linearly polarized probe beam propagating along $x$. The probe's polarization rotates after passing through the cell, due to the spin-dependent optical susceptibility of the atoms, and is centered on the D1 lines. A balanced photo-detection setup measures the ensemble mean spin, $\langle S_x(t)\rangle$ \cite{katz2015coherent}.

The spin precession around $B=0.43 \,\textrm{mG}$ is measured for various powers of the optical protection beam $P_{\textrm{op}}$. In Fig.~\ref{fig:spectrum}a, the spectral content of the recorded signals is shown as the absolute value of the Fast Fourier Transform. At $P_{\textrm{op}}=0$, the spectrum exhibits a single peak, reflecting the indistinguishable Larmor precession of the two hyperfine manifolds at the same absolute frequency. Increasing $P_{\textrm{op}}$ separates the two precession frequencies $\omega_{\textrm{a}}=\delta_{\textrm{a}}(P_{\textrm{op}})+\omega_B$ and $\omega_{\textrm{b}}=\delta_{\textrm{b}}(P_{\textrm{op}})-\omega_B$, as highlighted by the orange and red lines. When these frequencies converge ($\omega_{\textrm{a}}=\omega_{\textrm{b}}=\omega$), near $P_{\textrm{op}}=9.7 \textrm{mW}$, the resonance condition is met, leading to a striking enhancement in the spectral amplitude due to significant spectral narrowing compared to lower or zero protection beam power. Notably, the presented data is unnormalized across measurements; all amplitudes were scaled by a single common factor.   

\begin{figure*}[t]
\begin{centering}
\includegraphics[width=16cm]{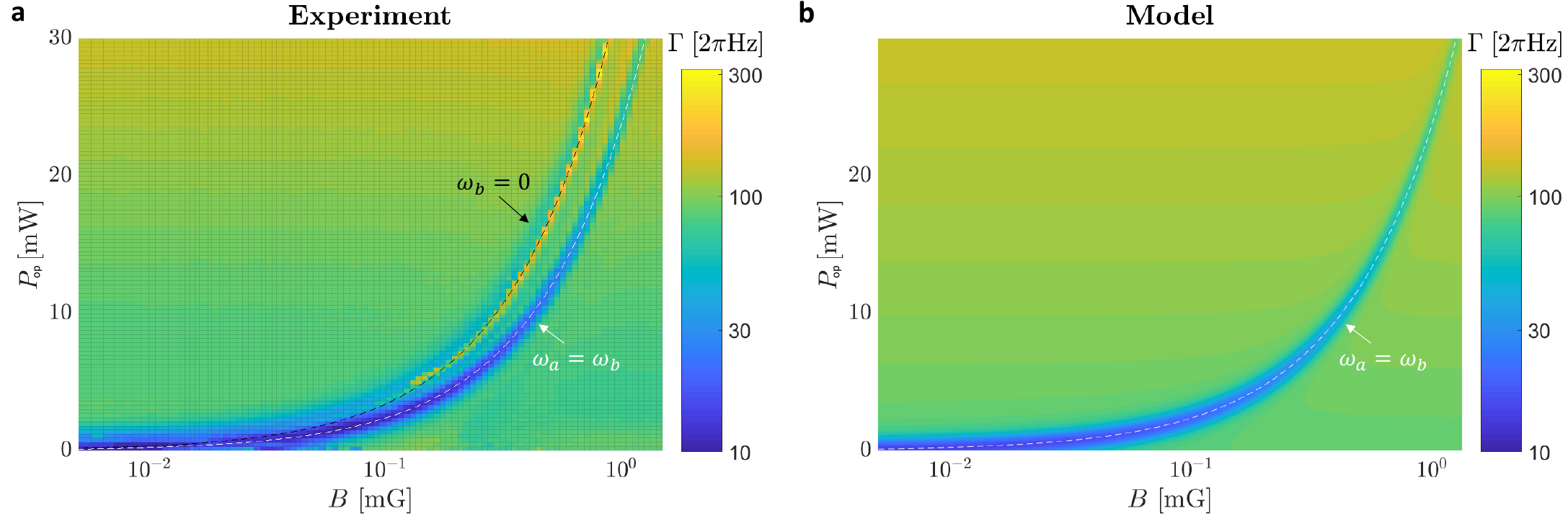}
\par\end{centering}
\centering{}\caption{\textbf{Protection from spin relaxation.} \textbf{a}, Measured fundamental relaxation rate $\Gamma$ of cesium vapor as a function of the magnetic field $B$ and power $P_{\textrm{op}}$ of the optical protection beam. The relaxation is minimized along the resonance line $\omega_{\textrm{a}}(B,P_{\textrm{op}})=\omega_{\textrm{b}}(B,P_{\textrm{op}})$ (white dashed line), providing up to an eightfold reduction in relaxation compared to precession without light shifts. \textbf{b}, Calculated fundamental relaxation rate using the hyperfine-Bloch model (see text and \cite{SI}). Increased relaxation is experimentally observed  for $\omega_{\textrm{b}}(B,P_{\textrm{op}})=0$ (black dashed line).\label{fig:GAMMA}}
\end{figure*}

In Fig.~\ref{fig:spectrum}b, we present time-domain signals recorded without the protection beam ($P_{\textrm{op}}=0$, top), and near resonance at $P_{\textrm{op}}=9.7\,\textrm{mW}$ (bottom), corresponding to the black and purple arrows in Fig~\ref{fig:spectrum}a. Fitting each signal to a single exponentially decaying sinusoid reveals a five-fold reduction in the decoherence rate near resonance compared to the case without the protection beam, limited only by the finite spin lifetime $T_1$ \cite{SI}. These results clearly demonstrate the suppression of spin-relaxation under synchronized precession of the two magnetic moments. 

We further study the dependence of the relaxation on the $B$ and $P_{\textrm{op}}$ across $10^4$ different experimental configurations. The measured signals are fit to a sum of two exponentially decaying sinusoids $\sum_{i=1}^2{A_ie^{-\gamma_it}\cos(\omega_it+\theta_i)}$ and the minimum relaxation rate, $\Gamma=\min(\gamma_1,\gamma_2)$, is shown in Fig.~\ref{fig:GAMMA}a. Evidently, a valley of low relaxation rates $\Gamma$ emerges along magnetic fields satisfying the resonance condition $\omega_{\textrm{a}}=\omega_{\textrm{b}}$, indicated by the white dashed line. This highlights decoherence suppression at magnetic fields significantly higher than those in the SERF regime at $P_{\textrm{op}}=0$. In \cite{SI}, we present two signals at the same magnetic field, demonstrating a suppression of decoherence by approximately ninefold. The relaxation rate shows a weak increase with $P_{\textrm{op}}$ and remains independent of the magnetic field, consistent with measured residual absorption of the protection beam. An increase in relaxation is also observed along the curve where $\omega_\textrm{b}(B,P_{\textrm{op}})=0$ (black dashed line).

We model the observed phenomena using the hyperfine-Bloch equations, valid in the low-polarization regime \cite{happer1977effect, xiao2021atomic, mouloudakis2022effects, katz2015coherent}, and account for additional relaxation processes detailed in \cite{SI}. From measurements, we estimate relaxation rates due to spin exchange as $R_{\textrm{se}}=(170\pm30)\,\textrm{s}^{-1}$, electron spin destruction as $R_{\textrm{sr}}=(85 \pm 15) \,\textrm{s}^{-1}$ and total-spin destruction as $R_{\textrm{u}}=(10 \pm 3) \,\textrm{s}^{-1}$), the latter two caused by collisions with weakly depolarizing paraffin-coated walls. Additional relaxation due to off-resonant absorption of photons from the protection beam, $R_{\textrm{op}}$, is also modeled. Figure.~\ref{fig:GAMMA}b shows the numerically calculated minimal relaxation, which reproduces the main experimental features. This confirms that synchronization of Larmor precession via hyperfine-selective vector light shifts effectively protects magnetic moments from spin-exchange relaxation.

\begin{figure}[b]
\begin{centering}
\includegraphics[width=8.2cm]{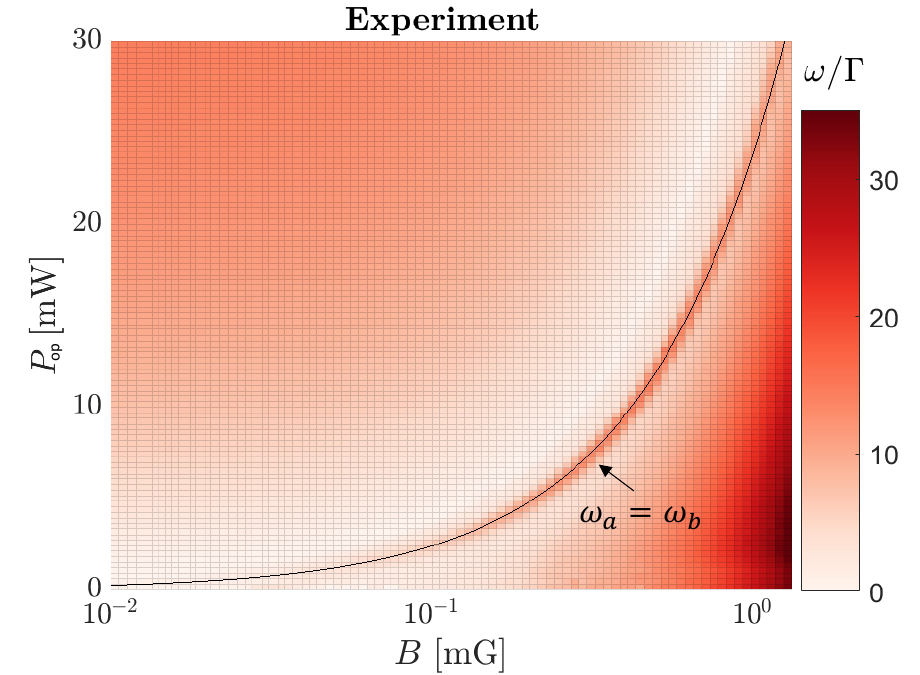}
\par\end{centering}
\centering{}\caption{\textbf{Experimentally measured number of visible precession cycles.} The number of visible spin-precession cycles, $Q={\omega}/\Gamma$, is considerably higher along the resonance line (black dashed line) compared with low magnetic fields ($Q\lesssim1$). This significant improvement in $Q$ is a unique feature of the optical protection technique. \label{fig:Qfactor}}
\end{figure}

We examine three distinct aspects of this technique compared to suppression methods without the protection beam. First, in the regime of frequent spin-exchange collisions $(R_{\textrm{se}}\gg R_{\textrm{sr}},R_{\textrm{u}})$ and low spin polarization ($P\ll1$), the zero-field relaxation rate for $I=7/2$ is $\Gamma_0=R_{\textrm{u}}+\frac{1}{22} R_{\textrm{sr}}$ \cite{SI}.
It increases with frequency mismatch as $\Gamma\approx\Gamma_0+0.3|\omega_{\textrm{a}}-\omega_{\textrm{b}}|^2/R_{\textrm{se}}$ for $|\omega_{\textrm{a}}-\omega_{\textrm{b}}|\ll R_{\textrm{se}}$. Without the protection beam, $\omega_{{a}}=-\omega_{\textrm{b}}=\omega_{B}$, causing relaxation to grow quadratically with magnetic field amplitude at low fields and become field-independent but linear in $R_{\textrm{se}}$ at high fields \cite{SI}. With the protection beam, relaxation can be suppressed even at higher magnetic fields, with the achievable field amplitude practically limited by the maximum attainable vector light shift.

Second, the relationship between the precession rate $\omega$ and relaxation rate $\Gamma$ highlights another key advantage. Without the protection beam, dynamics in the low-field regime are relaxation-dominated, with $Q=\omega/\Gamma<1$ (where ${\omega}=\tfrac{4}{11}\omega_{B}$). By contrast, the protection beam decouples $\Gamma$ from $\omega$, enabling $Q\gg1$. Figure.~\ref{fig:Qfactor} shows the experimentally measured $Q$ factor, with the black line marking the resonance condition $\omega_{\textrm{a}}=\omega_{\textrm{b}}$. Along this line, the number of visible oscillations is significantly higher than without the beam, due to suppressed relaxation and higher oscillation frequencies. In \cite{SI}, we provide an example demonstrating an order of magnitude improvement in the $Q$ factor for two measured signals with similar precession frequencies and numerically show that further enhancements can be achieved by optimizing the protection beam parameters.

Finally, we extend our analysis to numerical simulations beyond the low-polarization limit \cite{SI}. In the SERF regime without the protection beam, the precession frequency strongly depends on spin polarization $P$. For cesium, the effective gyromagnetic ratio changes by a factor of $2.75$ between $P=0$ and $P=1$. By contrast, our technique ensures that the precession rate is entirely independent of $P$, as shown in \cite{SI}, allowing ensembles at different spin temperatures to precess uniformly, independent of the polarization which often depend on the parameters of the optical pumping beam.

In summary, we have studied the spin-relaxation of a warm spin-gas in the presence of a magnetic field and demonstrate theoretically and experimentally a technique to suppress spin-relaxation at a range of parameters that is distinct from other techniques including the range of magnetic fields, the quality factor, and the independence of the gyromagnetic ratio on the spin polarization.

The technique relies on achieving uniform spin precession across all atoms. Since synchronization is implemented via vector light shifts that depend on the local beam intensity, imperfections such as beam attenuation and finite transverse size could, in principle, disrupt the synchronization condition. Remarkably, the rapid thermal motion of atoms within the cell leads to motional narrowing, which averages out these spatial inhomogeneities and renders the non-uniformity negligible over a broad range of parameters \cite{SI}. Moreover, we show that the requirements on the beam's frequency and intensity stability are substantially relaxed.

The technique is particularly advantageous for systems utilizing anti-relaxation-coated cells with well-resolved optical spectra. As detailed in \cite{SI}, it has potential applications in quantum information systems (e.g., \cite{julsgaard2004experimental,julsgaard2001experimental,katz2020long,moller2017quantum,guarrera2019parametric,hammerer2010quantum,bao2020spin,qu2020sub}) that benefit from relaxation suppression while requiring high-frequency precession to mitigate low-frequency noise. It may also support sensing applications, such as all-optical magnetometers (e.g., \cite{patton2014all,ding2023dual}) operating in environments with a known or slowly varying magnetic field direction, as the protection mechanism assumes that the optical axis is aligned with a stable field orientation. In such regimes, the protection mechanism can improve coherence, enhance accuracy, and approach the fundamental sensitivity of low-field SERF magnetometers \cite{allred2002high,kominis2003subfemtotesla}. Extensions to closed-loop feedback control could enable dynamic tracking of field variations and further expand the method's applicability.

\begin{acknowledgments}
We thank Roy Shaham, Ohad Yogev, Constantine Feinberg, Gil Ronen, Yaron Artzi and Tal David for fruitful discussions. 
\end{acknowledgments}

\bibliography{Refs}

\end{document}